\documentclass{aa}
\begin{document}

      \title{Self-Similar Evolutionary Solutions of Self-Gravitating, Polytropic $\beta$-Viscous Disks}


      \author{Shahram Abbassi\inst{1} \and Jamshid Ghanbari\inst{2,3} \and Fatemeh Salehi\inst{2,4}
       }

      \offprints{S. Abbassi, \email{sabbassi@dubs.ac.ir}}
      \institute{Department of Physics, Damghan University of Basic
          Sciences,
                Damghan, Iran
       \and
            Department of Physics, School of Sciences, Ferdowsi
                  University of Mashhad, Mashhad, Iran
       \and
                 Department of Physics and Astronomy, San Francisco State
              University, 1600 Holloway, San Francisco, CA 94132
       \and
                 Department of Physics, Khayam Institute of Higher
                 Education, Mashhad, Iran
             }
\date{Received ... / Accepted }
\abstract {} { We carry out the effect of $\beta$-prescription for
viscosity which introduced by Duschel et al. 2000 \& Hure, Richard
\& Zhan 2001, in a standard self-gravitating thin disks. We were
predicted in a self-gravitating thin disk the $\beta$-model will
have different dynamical behavior compare the well known
$\alpha$-prescriptions.}{ We used self-similar methods for solving
the integrated equations which govern the dynamical behavior of
the thin disks.} {We present the results of self-similar solutions
of the time evolution of axisymmetric, polytropic,
self-gravitating viscous disks around a new born central object.
We apply a $\beta$-viscosity prescription which has been derived
from rotating shear flow experiments ($\nu=\beta r^2\Omega$).
Using reduced equations in a slow accretion limit, we demonstrate
inside-out self-similar solutions after core formation in the
center. Some physical quantities for $\beta$-disks are determined
numerically.We have compared our results with $\alpha$-disks under
the same initial conditions. It has been found that the accretion
rate onto the central object for $\beta$-disks more than
$\alpha$-disks at least in the outer regions where $\beta$-disks
are more efficient. Our results show that Toomre instability
parameter is less than one everywhere on the $\beta$-disk which
means that in such disks gravitational instabilities can be
occurred, so the $\beta$-disk model can be a good candidate for
the origin of planetary systems. Our results show that the
$\beta$-disks will decouple in the outer part of the disk where
the self-gravity plays an important role which is in agreement
with Duschl predictions. } {}


\maketitle

\section{Introduction}
\label{sec:intro}

Accretion disks are recognized as one of the objects that are
found around many astrophysical objects, such as active galactic
nuclei AGN, binary stars and young stellar objects. On the
observational side, evidence for disks in young stellar objects
gleaned both spectroscopically and through direct imaging is now
quite compelling (Beckwith $\&$ Sargent 1993, Storm et al, 1993).
Up to half of the solar type, pre-main sequence stars are
surrounding by diks of gas and dust, many of these disks having
masses similar to that expected for the early solar nebula
(Chandler, C. J. 1998). The accretion disks around pre-main
sequence stars are good candidates for the creation of planetary
systems. The structure of such disks is a subject of great
interest and has been studied both through self-similar solutions
assuming unsteady state (Mineshige $\&$ Umemura 1997; Mineshige et
al. 1997; Tsuribe 1999) and through direct numerical
hydrodynamical simulations (Igumenshchev $\&$ Abramowicz 1999;
Stone et al. 1999; Torkelsson et al. 2000). It is understood that
the most crucial factors are self-gravity and viscosity which have
great role on angular momentum transportation on the gas disk.
Accretion takes place because of the action of some form of
dissipation which releases the free energy of the shear flow as
heat, and so allows the disk material to fall deeper into the
potential well of the central object. In a simple picture
,Lynden-Bell and Pringle (1974) indicated that the dissipative
processes must take the form of a stress which transport angular
momentum outwards. It plays a significant role in many such
systems, ranging from protostellar disks to active galactic nuclei
(AGN). The self-gravity will modify the radial and vertical
equations and so can treat the dynamical behavior of the accretion
disks. In the standard thin accretion disk model, the effect of
self-gravity is neglected, and only pressure supports the vertical
structure. By contrast, the theory of self-gravitating accretion
disks is less developed and in the traditional model of accretion
disks, the self-gravity is ignored just for simplicity (e.g.
Pringle 1981), although the self-gravity can describe the
deviation from Keplerian rotation velocity in some AGN and flat
infrared spectrum of some T Tauri stars. From the observational
point of view, there are already some clues that the disk
self-gravity can be important both in the context of proto-stellar
disks and in the accretion disks around super massive black holes
in the AGN. However, the detailed comparison with observations is
limited by the lack of detailed models of self-gravitating disks
and by an incomplete understanding of the basic physical processes
involved.

The study of self-gravity in a most general case is difficult and
because of these complexities, authors usually study the effects
related to the self-gravity either in the vertical structure of
the disk (e.g. Bardou et al. 1998) or in the radial direction
(e.g. Bodo $\&$ Curir 1992). Disks in the AGN are thought to be
relatively light in the sense that the ratio of $\frac{M_d}{M_*}$
is around a few percents (where $M$ and $M_*$ are the masses of
the the disk and central star). Usually self-gravity plays
\textbf{its} role at large distances from the central objects
(Shlosman $\&$ Begelman 1987), and mainly in the direction
perpendicular to the plane of the disk. But in the accretion disks
around young stellar objects or pre-main-sequence stars,
self-gravity can be important in all parts of the disk in both
vertical and radial directions. Early numerical works of
self-gravitating accretion disks began with N-body modelling
(Cassen $\&$ Moosman 1981 ; Tomley, Cassen $\&$ Steinman-Cameron
1991). Shlosman $\&$ Begelman (1987) investigated the role of
self-gravity in AGN. Recently, Ghanbari and Abbassi (2004)
introduced a toy model which shows that at least in equilibrium of
a thick accretion disk, self-gravity is an important effect.

One of the basic concepts of the theoretical descriptions of
accretion disks and these dynamics, is the detailed knowledge
underlying the physics of viscosity in the disks. Because of all
detailed modeling of the structure and evolution of accretion
disks depending on the viscosity and its dependence on the
physical parameters, choosing the best viscosity model is quite
important. There is a belief that, the molecular viscosity is
inadequate to describe the observational evidence of some luminous
accretion disks so some kinds of turbulence viscosity are
required. Most investigators adopt so-called $\alpha$-model
introduced by Shakura (1972) and Shakura $\&$ Sanyeav (1973) that
gives the viscosity as the product of pressure scale height in the
disk ($h$), the velocity of the sound ($c_s$) and a parameter
$\alpha$ which contains all unknown physics. The models for the
structure and evolution of accretion disks in close binary systems
(e.g. dwarf novae and symbiotic stars) show that Shakura and
Sunyeav's parametrization with a constant $\alpha$ leads to
results that reproduce the overall observed behavior of the disks
quite well. And it has been recently recognized that accretion
disks treated by a week magnetic field are subjected to MHD
instabilities (Balbus $\&$ Hawly 1991), that can induce some kind
of turbulence in the disk, thereby being able to transport angular
momentum and to promote accretion processes. However in many
astrophysically interesting cases, such as the outer part of
proto-stellar disks, the ionization level is expected to be low,
reducing significantly the effect of magnetic field in the
dynamical behaviour of the disk. The realization that molecular
transport of angular momentum is so inefficient led the
theoreticians to look for another mechanism of transport of
angular momentum in accretion disks. The alternative mechanism
that would be a good candidate for transporting angular momentum
in accretion disks is any kind of turbulence. Actually the
$\alpha$-prescription is based on a kind of turbulence viscosity
but it has not any physical base to drive the origin of turbulence
in the model. On the other hand, laboratory experiments of
Taylor-Couette systems seem to indicate that, although Coriolis
force delays the onset of turbulence, the flow is ultimately
unstable to turbulence for Reynolds numbers larger than a few
thousands (e.g., see , Richard, Zahn, 1999 and Hure, Richard, Zahn
2001). Since in all kind of self-gravitating disks the Reynolds
number is extremely high, it was thought that the hydrodynamical
driven turbulent viscosity based on critical Reynolds number has
probably significant role in the distribution of angular momentum
in the accretion disks. The resulting turbulence would then
transport angular momentum efficiently. Recently, Duschl,
Stirittmatter and Biermann (2000) have proposed a generalized
accretion disk viscosity prescription based on the
hydrodynamically driven turbulence at the critical effective
Reynolds number,$\beta$-model , which is applied for both
self-gravitating and non self-gravitating disks and is shown to
yield the standard $\alpha$-model in the case of the shock
dissipation limited, non self-gravitating disks. They have shown
that in the case of fully self-gravitating disks this model may
explain the observed spectra of proto-planetary disks and yield a
natural explanation for the radial motions from the observed
metallicity gradients in the disk galaxy.

The basic equilibrium and dynamical structure of accretion disks
are now well understood, as long as the standard model based on
the $\alpha$-viscosity prescription (Shakura $\&$ Sunyeav 1973) is
believed. Nevertheless, it is not easy to follow its dynamical
evolution, mainly because the basic equations of the system are
highly non-linear, specially when the system is self-gravitating
(Paczynski 1978; Fukue $\&$ Sakamoto 1992). To follow the
non-linear evolution of dynamically evolving systems, in general,
the technique of the self-similarity is sometimes useful.
Self-similar assumptions enable us to simplify the governing
equations. Self-similar solutions have a wide range of
applications in astrophysics. Several classes of self-similar
solutions were known previously (Pringle 1974, Filipov 1984), but
all of them considered the disk in a fixed, central potential. But
a class of self-similar solution had been provided during last
years that contained self-gravity (Mineshige $\&$ Umemura 1997).
They had found a self-similar solution for a time evolution of
isothermal, self-gravitating $\alpha$-viscous disk. This solution
describes a homologous collapse of a disk via self-gravity and
viscosity. They found that the disk structure and evolution are
distinct in the inner and outer parts. The effect of self-gravity
in the collapse of polytropic self-gravitating viscous disk has
been investigated by Mineshige, Nakayama and Umemura (1997).

Following the Duschl et al. (2000) suggestion for
$\beta$-prescription for viscosity, we apply this model for a thin
self-gravitating disk around newborn stars. At first, it may seem
that using other forms of viscosity is not an important issue,
because just one should change the mathematical forms of the
equations. But this simplified appearance did not force authors
from such studies because of possible interesting results that one
may obtain from other viscosity prescriptions. All these lead us
to explore a self-gravitating disk using other viscosity
prescriptions. However, all viscosity prescriptions are suffering
from having phenomenological backgrounds rather than physically
confirmed backgrounds. We think while we don't have a clear
picture of the turbulence in disks, all such prescriptions are
standing on the same level as for their physical backgrounds. On
the other hand, self-gravitation in a disk is a highly nonlinear
process as a result of the complex behaviors of the various
physical agents of the system, in which the turbulent viscosity
and its prescriptions have a vital role. Thus, one may naturally
ask what would happen if other forms of viscosity prescriptions in
a self-gravitating disk is being used. When we searched in the
literature, we found the $\beta$-prescription as only
experimentally tested viscosity prescription. Indeed, there are
few studies about $\beta$-prescription comparing the
$\alpha$-model, clearly because of its newly proposed form. One
may refer to many papers to see physical considerations and
applications of the $\beta$-model: Mayer \& Duschl (2005); Weigelt
et al. (2004); Pott et al. (2004); Granato et al. (2004); Mathis,
Palacios \& Zahn (2004); Richard \& Davis (2004). We expect to
find different dynamical behaviors and we will show that in these
disks the gravitational fragmentation can take place everywhere in
the disks. So it will be a good description for the formation of a
proto-planetary disk.

\section{Formulation of Equations for Self-Similar Variables}
Self-similar behavior provides a set of unsteady solutions to the
self-gravitating fluid equations. On the other hand, many physical
problems often attain self-similar limits for a wide range of
initial conditions. Also self-similar properties allow us to
investigate properties of the solutions in arbitrary detail,
without any of the associated difficulties of numerical
hydrodynamics.\\

\subsection{The Basic Equations}
In order to study the accretion processes of a thin disk under the
effect of the self-gravity and viscosity, we consider axisymmetric
polytropic disks using the cylindrical coordinates ($r,\phi,z$).
We assume that the accretion disks are geometrically thin in the
vertical direction and symmetric in the azimuthal direction. The
model is described by the fundamental governing equations
which are written as follow :\\

\begin{equation}
\\\frac{\partial \rho}{\partial
t}+\frac{1}{r^2}\frac{\partial}{\partial r}(r^2\rho
v_r)=0\label{eqcon}
\end{equation}
\begin{equation}
\\\frac{\partial v_r}{\partial t}+v_r \frac{\partial v_r}{\partial
r}-\frac{v_{\phi}^2}{r}=-\frac{1}{\rho}\frac{\partial p}{\partial
r}-\frac{\partial \Phi}{\partial r}\label{eqeul}
\end{equation}
\begin{equation}
\\\rho\frac{\partial (rv_{\phi})}{\partial t}+\rho v_r\frac{\partial (rv_{\phi})}{\partial
r}=\frac{1}{r^2}\frac{\partial}{\partial r}(\nu\rho r^4
\frac{\partial\Omega}{\partial r})\label{eqmom}
\end{equation}

where $\rho$, $p$, $v_r$ and $v_{\phi}$ are density, pressure,
radial and azimuthal components of velocity of the gas disk and
also $\Phi$ is the gravitational potential of the gas disk inside
of the radius r. We assume a polytropic relation between the gas
pressure and density :

\begin{equation}
\\P=K\rho^{\gamma}
\end{equation}
with K and $\gamma$ being constants. The polytropic index $\gamma$
describes the adiabatic pressure-density relation. In subsequent
analysis, we vary it and represent its effect on some physical
quantities. The vertical extent of the disk at any radius is given
by h, the half thickness of the disk:

\begin{equation}
\\h=\frac{c_s}{(4\pi G\rho)^{\frac{1}{2}}}=\frac{c_s^2}{2\pi G \sigma}
\end{equation}
where $c_s$ is the sound speed and $\sigma$ is the surface
density.

The solution of these equations, give us the dynamical evolution
of the disk which strongly depend on the viscosity model. So the
study of dynamical behavior of the accretion disks is postponed to
the more information about the viscosity.

\subsection{Nondimensionalization}
Before we actually begin solving equations
(\ref{eqcon})-(\ref{eqmom}), it is convenient to nondimensionalize
the equations. The essence of self-similar model is the existence
of only two dimensional parameters in the problem, viz., $K$ and
the gravitational constant G. It is assumed, and this is born out
by numerical calculations, that any additional parameters, such as
the initial central density affects only transients, and theis
memory is quickly lost, at least in the part of the flow in which
the density greatly exceeds the initial central density. If that
is the case, then only one dimensionless combination of radius r
and time t can be found
(Mineshige et al. 1997, Yahil 1983):\\

\begin{equation}
\\\xi= K^{-\frac{1}{2}}G^{\frac{\gamma-1}{2}}r(t)^{\gamma-2}
\end{equation}

This determines the dimensionless parameterizations of any
similarity solution (Mineshige et al. 1997). Note that, we
consider only $t>0$ for our work and the origin of time $(t=0)$
corresponds to the core formation epoch. Hence we have:\\

\begin{equation}
\frac{\partial}{\partial t}=\frac{\partial}{\partial
t^\prime}+(\gamma-2)\frac{\xi}{t^\prime}\frac{\partial}{\partial
\xi}
\end{equation}
and
\begin{equation}
\frac{\partial}{\partial
r}=K^{-\frac{1}{2}}G^{\frac{\gamma-1}{2}}(t^\prime)^{\gamma-2}\frac{\partial}{\partial
\xi}
\end{equation}
for the transformation $(t,r)\rightarrow (t^\prime=t,\xi)$.
Self-similarity allows us to reduce the self-gravitating fluid
equations from partial differential equations into ordinary
differential equations.

For changing the variables to dimensionless form, we used K and G;
because we require that all of the time-dependent terms disappear
in the self-similar forms of the equations. Other physical
quantities (functions of t and r) are transformed into
self-similar ones (functions of only $\xi$) as:

\begin{equation}
v_r(t,r)=K^{\frac{1}{2}}G^{\frac{1-\gamma}{2}}(t)^{1-\gamma}V_r(\xi),
\end{equation}
\begin{equation}
v_{\phi}(t,r)=K^{\frac{1}{2}}G^{\frac{1-\gamma}{2}}(t)^{1-\gamma}V_{\phi}(\xi),
\end{equation}
\begin{equation}
j(t,r)= KG^{1-\gamma}(t)^{3-2\gamma}J(\xi),
\end{equation}
\begin{equation}
\sigma(t,r)=(2\pi)^{-1}K^{\frac{1}{2}}G^{-\frac{1+\gamma}{2}}(t)^{-\gamma}\Sigma(\xi),
\end{equation}
\begin{equation}
\rho(t,r)=(4\pi\gamma)^{-\frac{1}{\gamma}}K^{\frac{1}{2}}G^{-1}(t)^{-2}\Sigma^{\frac{2}{\gamma}}(\xi),
\end{equation}
\begin{equation}
p(t,r)=(4\pi\gamma)^{-1}KG^{-\gamma}(t)^{-2\gamma}\Sigma^2(\xi),
\end{equation}
\begin{equation}
\nu(t,r)= KG^{1-\gamma}(t)^{3-2\gamma}\nu^\prime(\xi)
\end{equation}
\begin{equation}
m(t,r)=K^{\frac{3}{2}}G^{\frac{1-3\gamma}{2}}(t)^{4-3\gamma}M(\xi).
\end{equation}
where
\begin{equation}
J=\xi V_\phi~~~,~~~~M=\frac{\xi\Sigma u}{3\gamma-4}.
\end{equation}
with respect to another form for the continuity equation :

\begin{equation}
\frac{\partial m}{\partial t}+ v_r\frac{\partial m}{\partial
r}=0~~~~~~,~~~~~m= 2\pi\int_0^r \sigma r dr
\end{equation}
thus, we have
\begin{equation}
\dot{M}=\frac{\xi\Sigma V_r}{3\gamma-4}
\end{equation}

In order to solve the equations, we need to assign the kinematic
coefficient of viscosity $\nu$. Although there are many
uncertainties about the exact form of viscosity, as we mentioned
in Introduction, we employ the $\beta$-prescription introduced by
Duschl et al.:
\begin{equation}
\\\nu^\prime=\beta \xi V_{\phi}
\end{equation}
where $\nu^\prime$ is on a dimensionless form. We expect when the
disk is non-self-gravitating, it leads to standard
$\alpha$-prescriptions. With substituting this prescription into
the above equations we can investigate the dynamical evolution of
the disk.

\subsection{Transformation of the basic Equations}

Substituting the above transformations in equations
(\ref{eqcon})-(\ref{eqmom}), we can introduce a set of coupled
ordinary differential equations. The basic equations are then
transformed into the following forms:

\begin{equation}
\frac{1}{\xi}\frac{d}{d\xi}(\xi\Sigma
u)=(3\gamma-4)\Sigma\label{eqdcon}
\end{equation}
\begin{equation}
u\frac{du}{d\xi}=-\frac{c^2}{\Sigma}\frac{d\Sigma}{d\xi}-\frac{M}{\xi^2}+
\frac{J^2}{\xi^3}+(2\gamma-3)u+(2-\gamma)(\gamma-1)\xi\label{eqdrmo}
\end{equation}
\begin{equation}
u\frac {dJ}{d\xi}=\frac{1}{\Sigma\xi}\frac{d}{d\xi}(\beta \xi^3
\Sigma J
\frac{d}{d\xi}(\frac{J}{\xi^2}))+(2\gamma-3)J\label{eqdzmo}
\end{equation}
where $u$ is defined as,

\begin{equation}
u=V_r -(2-\gamma)\xi\label{u}
\end{equation}
and $c^2$ is a constant:

\begin{equation}
c^2=2(4\pi\gamma)^{\frac{1-\gamma}{\gamma}}\Sigma^{\frac{2\gamma-2}{\gamma}}
\end{equation}

Now we have a set of complicated differential equations that must
be solved under appropriate boundary conditions. Although a full
numerical solutions to these equations would now be possible; it
is more instructive to proceed by analyzing the model in some
restrictive cases such as one on slow accretion limits.

\section{\textbf{Behavior of the Solutions}}
\subsection{\textbf{Slow Accretion Limit}}

We consider the fluid equations for a thin disk in the slow
accretion approximation. The most substantive aspect of the slow
accretion approximation consists of rotationally supported disks
when the viscous timescale is much larger than the dynamical
timescale. In addition, the pressure gradient force and the
acceleration term in this approximation are ignored. The slow
accretion approximation in disks has been used by Tsuribe (1999)
and Minshige et al. (1997) and many others.

In the slow accretion limit ($|V_r|\ll 1\ll V_\phi$), and in the
equation \ref{eqeul}, the Euler equation, we have radial
force-balance which means that only two terms on the right hand of
the equation balance each other so we have:
\begin{equation}
\frac{J^2}{\xi^3}-\frac{M}{\xi^2}=0
\end{equation}
hence we have:

\begin{equation}
J= (\xi M)^{\frac{1}{2}}=(\frac{\Sigma
u}{3\gamma-4})^{\frac{1}{2}}\xi\label{j1}
\end{equation}

We can make some simplifications in the equation of continuity in
self-similar form, then it becomes:

\begin{equation}
\frac{dln \Sigma}{dln \xi}=-1-\frac{dln u}{dln
\xi}+(\frac{3\gamma-4}{u})\xi\label{dsigma}
\end{equation}

Before driving other equations in a slow accretion limit, we
introduce the distribution of the initial specific angular
momentum, $j$, as (Tsuribe 1999):

\begin{equation}
j(r)=q{\frac{G}{c_s}}m(r)\label{j2}
\end{equation}
where $j$ and $m$ are self-similar angular momentum and the total
mass within the cylindrical radius r, respectively. The variable q
is a dimensionless quantity has a constant value for the invicid
equilibrium solutions (Mestel 1963; Toomre 1982) and variable
value when the viscosity effect is included. Thus in dimensionless
form,
\begin{equation}
J=qM\label{j3}
\end{equation}
Now using Eqs.(\ref{j1}) and (\ref{j3}) we find that:

\begin{equation}
\Sigma^{\frac{1}{2}}=\frac{(3\gamma-4)^{\frac{1}{2}}}{qu^{\frac{1}{2}}}\label{sigma}
\end{equation}

With Eqs. (\ref{eqdzmo}),(\ref{u}),(\ref{j1}) ,(\ref{dsigma})and
(\ref{sigma}) we can finally find a simple differential equation
for $V_r$:

\begin{displaymath}
\frac{dV_r}{d\xi}=\frac{q}{\beta}\frac{V_r[V_r\mp(2-\gamma)\xi]^2}{\xi[\pm3V_r\mp(3\gamma-2)\xi]}+\frac{[3(2-\gamma)\pm5(3\gamma-4)]V_r}
{[\pm3V_r\mp(3\gamma-2)\xi]}
\end{displaymath}
\begin{equation}
+\frac{[(2-\gamma)(-18\gamma+22)\mp2(3\gamma-4)^2]\xi}{[\pm3V_r\mp(3\gamma-2)\xi]}\label{v}
\end{equation}

 Now we can use the standard fourth-order Runge-Kutta scheme
to integrate this ordinary differential equation. We can also use
the asymptotic solutions of this equation near the origin and the
outer part of the disk as a boundary condition.

\subsection{Numerical Analysis}

To solve equation (\ref{v}) numerically, we need one boundary
condition. We derived asymptotic solutions in the limits
$\xi\rightarrow 0$ and $\xi\rightarrow \infty$ for $V_r$ where
these asymptotic values can be used as boundary condition.

\begin{equation}
V_r = \frac{5\gamma-6}{6\gamma-7}\{
\xi-\frac{q_0}{\beta}\frac{(\gamma-1)^2}{6\gamma-7}\xi^2\}~~~~~~~\xi\rightarrow0
\end{equation}
\begin{equation}
V_r=-\frac{\beta}{q_0}\frac{[(2-\gamma)(22-18\gamma)-2(3\gamma-4)^2]}{(2-\gamma)^2}~~~~\xi\rightarrow\infty
\end{equation}
where $q_0$ is the asymptotic value of q in the outer radius. The
equation (\ref{v}) is integrated in the limits of the following
equations (eqs. 33 and 34). It was found that only for some
special values of the physical parameter such as $q, \gamma,
\beta$ there exist physical solutions satisfied at both boundary
conditions. By solving this equation we will have  $V_r$ profile
as a function of $\xi$ and other profiles ($V_\phi, \Sigma
\dot{M}...)$ can be obtained easily. In figures 1 and 2, we show
radial and azimuthal velocity distributions for some $\gamma$ and
$\beta$ values, respectively. Figures 3 and 4 indicate surface
densities and mass accretion rates profiles for some $\gamma$ and
$\beta$ values. Also we compare $\beta$ and $\alpha$ disks for
$\gamma=1.1$ in all figures. The self-similar variables are
functions of $\xi$. The behavior of the solutions predicted by the
beta viscosity makes much more radial velocity (see Figure 1)
compared to the $\alpha$-model (at least in the outer part of the
disk where the self-gravity has an important role) and also for
the azimuthal component of self-similar velocity,
$\beta$-viscosity leads to faster velocities compared to the
$\alpha$-model (see Figure 2). Therefore, the viscosity has more
efficient role on the redistribution of angular momentum, and it
leads to more radial flow and accretion rates.

It is predicted that the outer parts of a thin, viscous disk
around QSOs, self-gravity has an important role. This effect is
investigated by Toomre parameter (Toomre 1964), such that the
local gravitational instability occurs where $Q<1$ and where $Q>1$
the disk is stable against the gravitational fragmentation. So it
is useful to calculate Q values to compare gravitational stability
in $\beta$ and $\alpha$ disks. The Toomre stability parameter for
an epicyclic motion is:
\begin{equation}
Q=\frac{c_s k}{\pi G\sigma}=\frac{2J\sqrt
2y}{\xi^2}K^{\frac{\gamma-1}{4}}(4\pi)^{\frac{1-\gamma}{2\gamma}}
\gamma^{\frac{1}{2\gamma}}\Sigma^{\frac{-1}{\gamma}}
\end{equation}
where
\begin{equation}
k=\Omega(4+2\frac{d\log\Omega}{d\log r})^\frac{1}{2}
\end{equation}
is the epicyclic frequency and $\frac{dJ}{d\xi}=\frac{J}{\xi}y$.
So if $\gamma=1$, we obtain $Q=\frac{2J\sqrt 2y}{\Sigma \xi^2}$
(Tsuribe 1999).

In figure 5, we show the distribution of the Toomre Q value for
some $\gamma$ and $\beta$ values. We also compare $\beta$ and
$\alpha$ disks for $\gamma=1.1$. In figure 6, the angular momentum
coefficient $q=\frac{J}{M}$ is plotted as a function of $\xi$ for
some $\gamma$ values at $\beta=10^{-3}$. Also we see its behavior
for $\beta$ and $\alpha$ disks.

\section{Time scales}
For estimating the effect of viscosity on the evolution of
accretion disks we can compare the viscose time scale with
dynamical time scale. The dynamical time scale $\tau_{dyn}$ is
given by:
\begin{equation}
\tau_{dyn}=\frac{1}{\Omega}
\end{equation}
where $\Omega$ in the non self-gravitating (NSG) and Keplerian
self-gravitating (KSG) is given by the mass of the central
accretor and by the radius. But in the fully self-gravitating
disks, $\Omega$ is determined by solving Poisson's equation. The
time scale of viscose evolution $\tau_{visc}$ is given by:
\begin{equation}
\tau_{visc}=\frac{r^2}{\nu}
\end{equation}
where in the standard non self-gravitating and geometrically thin
accretion disks where $h << r $ ( case $\alpha$-disks ), this
leads to
\begin{equation}
\tau_{visc}^{non-SG}=(\frac{r}{h})^2\frac{\tau_{dyn}}{\alpha}
\end{equation}
where in full self-gravitating (FSG) or KSG disks ( $\beta$-disks)
is given by
\begin{equation}
\tau_{visc}^{FSG}=\tau_{visc}^{SG}=
\tau_{visc}^{KSG}=\frac{\tau_{dyn}}{\beta}
\end{equation}

So with $\alpha < 1$ and $\beta << 1$ under all circumstances
$\tau_{visc} >> \tau_{dyn}$. So the slow accretion limit will be
confirmed. The best approximation for $\alpha$ parameter in the
standard accretion model is less than one, $\alpha\sim 0.1$ but in
the $\beta$ model which based on the critical Reynolds number beta
is approximately $\beta \sim 10^{-2}-10^{-3}$. ($\frac{h}{r}$) in
the outer part of accretion disks is too small, so in the outer
part, of the disk where self-gravity of the disk dominates the
beta viscous time scale is less than alpha viscous time scale.
Therefore, the dynamical evolution of $\beta$-disks at least in
the outer part is faster than the inner parts. Although, in the
case of non-SG disks $\beta$-model can recover the standard
$\alpha$-disks. So with $\beta$-model we can reconstruct a better
picture in the equilibrium of galactic disks and protoplanetary
disks. As we can see in the Fig.1, in comparison to alpha model,
the radial flow in the beta disk is quite high which relevant to
beta mechanism time scale. Because of the time scale of the beta
disk is less, so it can evolve disks effectively and it can
produce the large radial flow. In the case of galactic disks the
inflow velocities in the $\beta$-model suggests values in the
range $0.3-3 km s^{-1}$ which is quite difficult to measure
directly where $\alpha$-model suggest still lower values Duschl et
al. (2000). Many authors have suggested the radial abundance
gradients observed in our own and other disk galaxies maybe due to
radial motion and diffusive mixing associated with the turbulence
generated by eddy viscosity ( $\beta$-disks) (Lacey $\&$ Fall
1985, Sommer-Larsen $\&$ Yoshii 1990; Koppen 1994; Edmunds $\&$
Greenhow 1995). Such radial inflows are consistant with the
$\beta$-model, but could be described with other physical
processes such as the effect of magnetic field.

\input{epsf}
\begin{figure}
\centerline{{\epsfxsize=9cm\epsffile{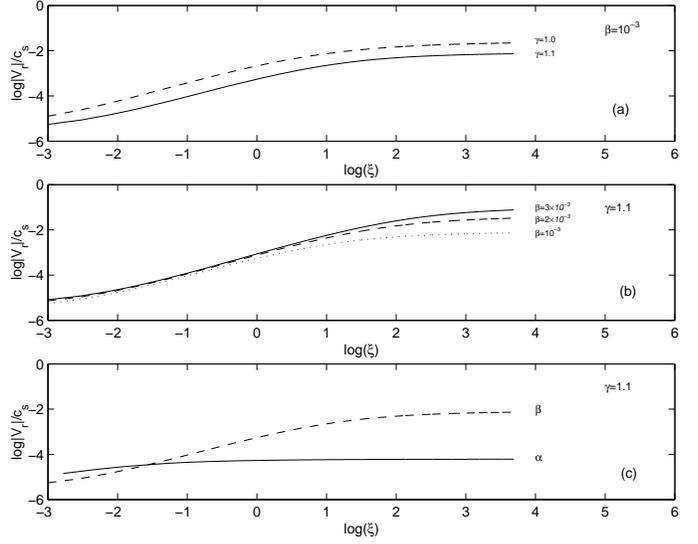}}} \caption{Radial
self-similar velocity distributions as a function
 of self-similar variable
 $\xi$ for a) $\gamma=1.0, 1.1$ at $\beta=10^{-3}$ b) $\beta=10^{-3}, 2\times10^{-3},
 3\times10^{-3}$ at $\gamma=1.1$ c) $\beta=10^{-3}$ and
 $\alpha=10^{-1}$ at $\gamma=1.1$.}
\end{figure}

\input{epsf}
\begin{figure}
\centerline{{\epsfxsize=9cm\epsffile{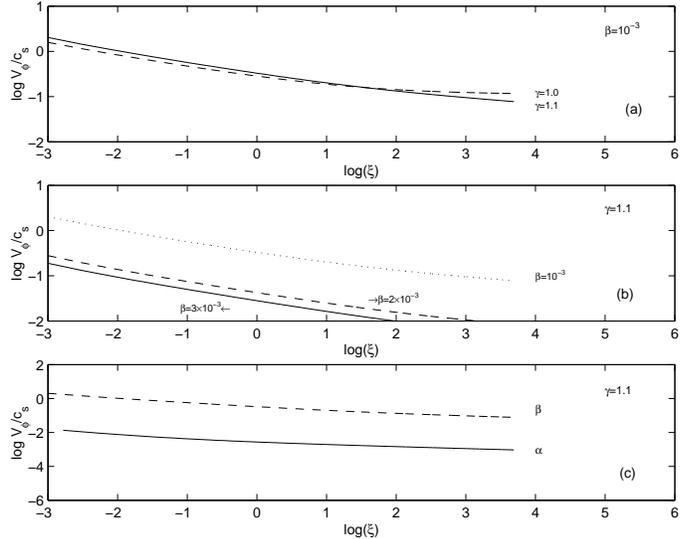}}} \caption{Azimuthal
self-similar velocity distributions as a function
 of self-similar variable
 $\xi$, corresponding to a) $\gamma=1.0, 1.1$ at $\beta=10^{-3}$ b) $\beta=10^{-3}, 2\times10^{-3},
 3\times10^{-3}$ at $\gamma=1.1$ c) $\beta=10^{-3}$ and
 $\alpha=10^{-1}$ at $\gamma=1.1$.}
\end{figure}

\input{epsf}
\begin{figure}
\centerline{{\epsfxsize=9cm\epsffile{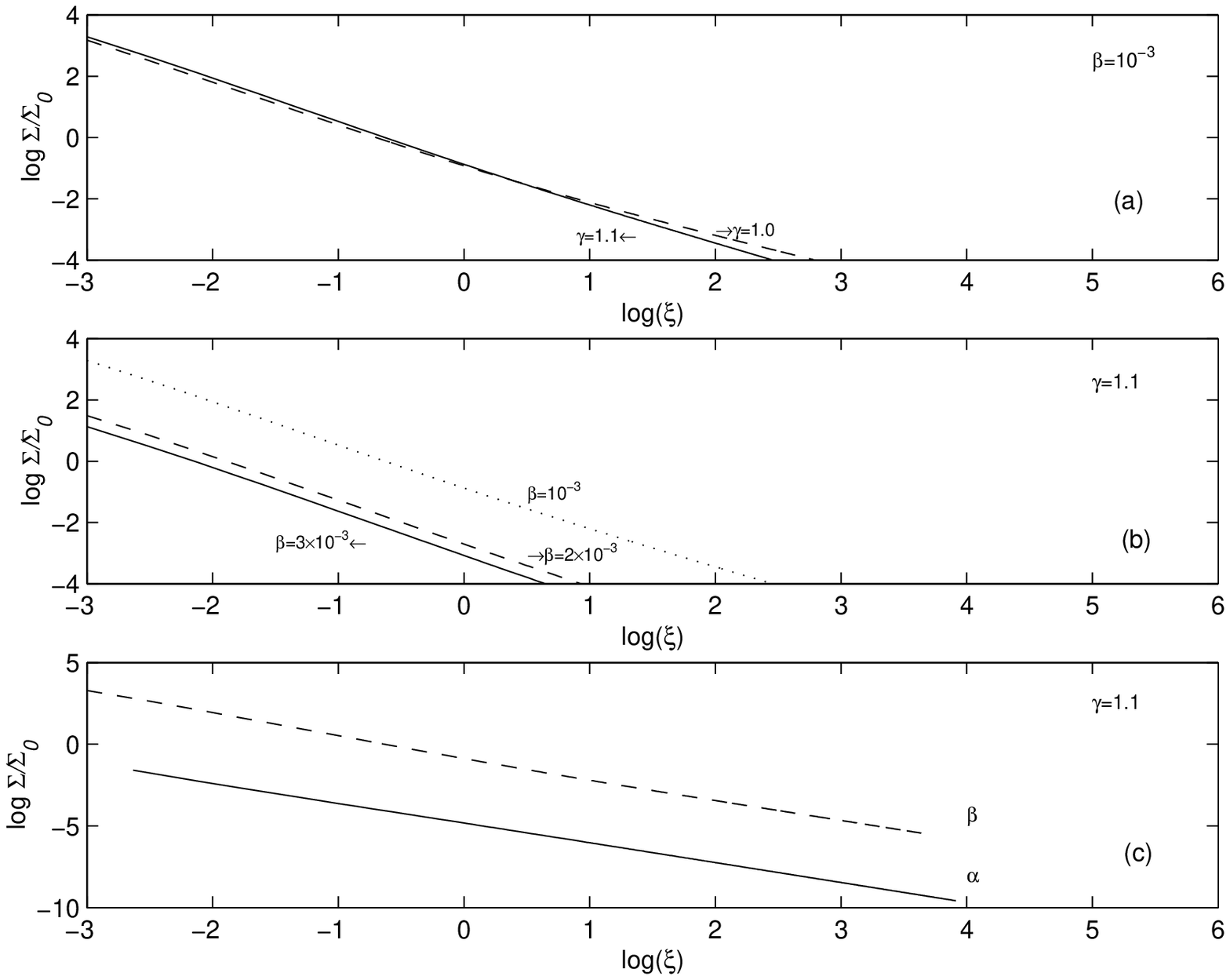}}}
\caption{Self-similar Surface densities distributions as a
function
 of self-similar variable
 $\xi$, corresponding to a) $\gamma=1.0, 1.1$ at $\beta=10^{-3}$ b) $\beta=10^{-3}, 2\times10^{-3},
 3\times10^{-3}$ at $\gamma=1.1$ c) $\beta=10^{-3}$ and
 $\alpha=10^{-1}$ at $\gamma=1.1$}
\end{figure}

\input{epsf}
\begin{figure}
\centerline{{\epsfxsize=9cm\epsffile{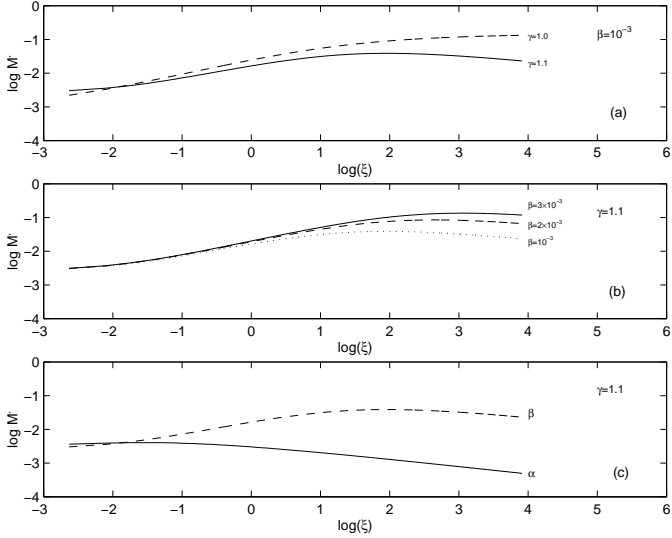}}}
\caption{Self-similar mass accretion rates distributions as a
function
 of self-similar variable
 $\xi$, corresponding to a) $\gamma=1.0, 1.1$ at $\beta=10^{-3}$ b) $\beta=10^{-3}, 2\times10^{-3},
 3\times10^{-3}$ at $\gamma=1.1$ c) $\beta=10^{-3}$ and
 $\alpha=10^{-1}$ at $\gamma=1.1$}
\end{figure}

\section{Concluding Remarks}

The $\beta$-prescription based on the assumption that the
effective Reynolds number of the turbulence dose not fall bellow
the critical Reynolds number. In this parametrization the
viscosity is proportional to the azimuthal velocity and the
radius. This model yields physically consistent models of both
Keplerian and fully self-gravitating accretion disks where in the
case of thin disks with sufficiently small mass, recover the
$\alpha$-disk solutions. Such $\beta$-disk models may be relevant
to protoplanetary accretion disks (Duschl et al. (2000)). In the
case of protoplanetary disks they yield spectra which are
considerably flatter than those due to non-self-gravitating disks,
in better agreement with observed spectra of these objects.

In this paper, we have considered the time-dependent evolution of
self-gravitating disks with $\beta$-prescription by self-similar
method for a thin, viscous disk. To do this, we started from
dimensionless basic fluid equations. In order to dominate gravity
and the centrifugal force, we consider the fluid equations for a
thin disk in the slow accretion approximation. It has been found
that evolution is described by solving a simple differential
equation (\ref{v}). We solved it numerically, beginning asymptotic
solution of this equation near the origin as a boundary condition.
Note should be taken that, we had the limitation to select
parameter $\gamma$ for essence of differential equations and the
fact that we seek just physical solutions.

\input{epsf}
\begin{figure}
\centerline{{\epsfxsize=9cm\epsffile{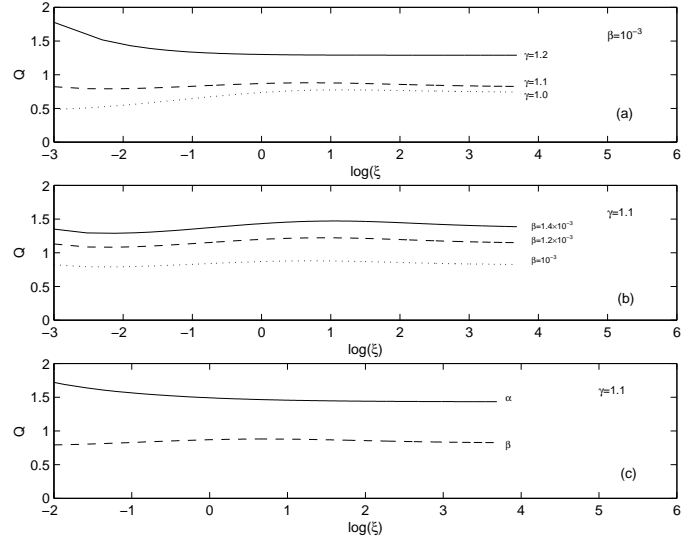}}}
\caption{Distribution of the Toomre Q value as a function of
self-similar variable
 $\xi$, corresponding to a) $\gamma=1.0, 1.1$ at $\beta=10^{-3}$ b) $\beta=10^{-3}, 2\times10^{-3},
 3\times10^{-3}$ at $\gamma=1.1$ c) $\beta=10^{-3}$ and
 $\alpha=10^{-1}$ at $\gamma=1.1$.}
\end{figure}

\input{epsf}
\begin{figure}
\centering \centerline{{\epsfxsize=9cm\epsffile{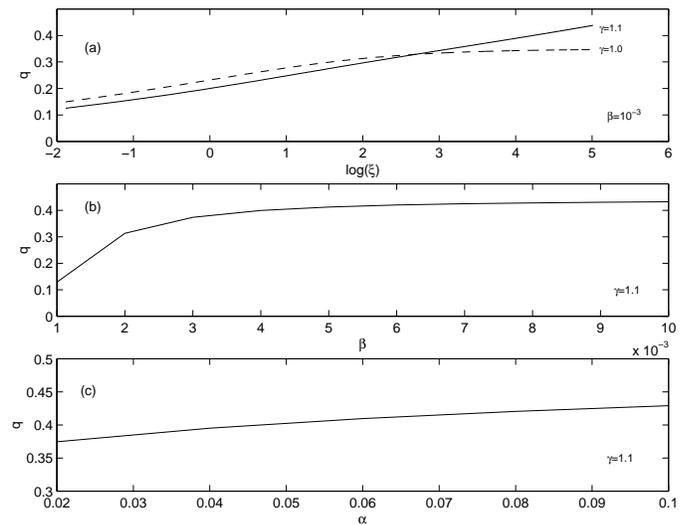}}}
  \caption{The angular momentum
coefficient profile, q, as a function self-similar variable $\xi$
corresponding to a) $\gamma=1.0, 1.1$ at $\beta=10^{-3}$. We see
its behavior versus the dimensionless viscosity parameter for b)
$\beta$ disks c) $\alpha$ disks at $\gamma=1.1$.}
 \label{Fig.6}
\end{figure}

The presented method shows that an increase of $\beta$ value
causes the azimuthal velocity to decrease but its general
distribution function doesn't vary throughout the disk. Also,
azimuthal velocities in $\beta$-disks are far more than
$\alpha$-disks (see Figure 2). So we expect the $\beta$-disks
evolve in different ways with respect to the $\alpha$-disks.

According to Figure 4, $\beta$-disks have larger mass accretion
rates than $\alpha$-disks. So, observably, we expect them to be
brighter than $\alpha$-disks. Also, we note that with the increase
of the $\beta$ value, $\dot M$ increases. Then mass flows increase
onto the central object. This follows more radial velocity and
less surface density (see Figures 1,3).

Comparing to $\alpha$-disks, q distribution ($q=\frac{J}{M}$)
seems non-smoothly. The q values are very small in the innermost
regions (see Figure 6). Whereas it is almost constant in the outer
regions. It seems in the outer part of the disk where the beta
viscosity is more efficient, the angular momentum is proportional
to the disk's mass inside the radius $r$. In order to study the
effect of self-gravity of thin $\beta$-disks, we plot the Toomre
parameter as a function of $\xi$. It is obvious that the
gravitational instabilities in $\beta$-disks are more pronounced
than $\alpha$-disks, $Q<1$. In Figure 5, Toomre parameter profile
can reveal this subject. So it can be expected that the
$\beta$-disk is a good model to describe planet formation around
new-born stars. In a global overview, we showed that in the outer
parts of the disk there is a difference between $\alpha$ and
$\beta$ models. These results were predicated by Duschl et. al
(2000).

Further, in order to study actual model and make a realistic
picture for a thin self-gravitating disks, one must investigate
energy exchange of the disk with its environment. In this case,
one should find a mechanism for transferring the thermal energy
from disk to outside environment; so we should add energy equation
to our model. Both $\alpha$ and $\beta$ models are
phenomenological prescriptions for disk viscosity. In an actual
model of viscosity, it is possible to combine these two models and
establish an exact description for different regions of the disk.
Also in real accretion disks, there are many important processes
other than viscosity that we have neglected. For example,
non-axisymmetric waves which are also expected to transport
angular momentum outward. Magnetic field and its influence are
neglected and sometimes magnetic braking is another possibility
for transporting angular momentum. However, these preliminary
solutions can be the beginning of our understanding deployment
from the physics governing the accretion disk.

\end{document}